# UMEX Viewer – Software Suite for High-Speed AFM Data Analysis and Its Applications

Kenichi Umeda[1,2*], Takahiro Watanabe-Nakayama[1], Yasuto Murayama[3,4], Hiroki Konno[1], Mikihiro Shibata[1], and Noriyuki Kodera[1*]

[1] *Nano Life Science Institute (WPI-NanoLSI),*

*Kanazawa University, Kakuma-machi, Kanazawa, Ishikawa, 920-1192, Japan.*

[2] *PRESTO/JST, 4-1-8 Honcho, Kawaguchi, Saitama 332-0012, Japan.*

[3] Department of Chromosome Science, National Institute of Genetics

[4] Department of Genetics, Graduate University for Advanced Studies (SOKENDAI),

Mishima, Shizuoka, 411-8540, Japan.

Corresponding Authors

Dr. Kenichi Umeda (E-mail: umeda.k@staff.kanazawa-u.ac.jp)

Prof. Noriyuki Kodera (E-mail: nkodera@staff.kanazawa-u.ac.jp)

## Abstract

Recent technical advances in atomic force microscopy (AFM) have led to the development of high-speed AFM (HS-AFM), which enables video-rate imaging and direct visualization of the nanoscale dynamics of biomolecules functioning in solution. Because HS-AFM rapidly generates large datasets, dedicated software is essential for efficient processing, organization, and analysis. Recent improvements in imaging speed have further increased data throughput, underscoring the need for faster analysis software with improved usability. To address this need, we developed Ultrafast Microscopy Exploration (UMEX) Viewer, a software suite for HS-AFM data analysis. UMEX Viewer provides an integrated environment for HS-AFM data handling, including image processing, drift correction, image and movie export, data management, and quantitative analysis. These capabilities have enabled the software to be applied to a wide range of recent studies. In this article, we describe the core functions of UMEX Viewer and demonstrate its practical applications for HS-AFM data analysis aimed at elucidating biological functions.

## 1. Introduction

Atomic force microscopy (AFM) is a technique that images surface structures by scanning a sample with a nanometer-scale sharp probe and has been widely used for the structural characterization of various materials, including biological specimens.[1-4] Along with advances in AFM instrumentation, numerous software packages for AFM data analysis have been developed.[5-16]

However, conventional AFM typically requires several minutes to acquire a single image, making it difficult to visualize the dynamics of biologically active molecules. To overcome this limitation, high-speed AFM (HS-AFM) was developed,[17-19] enabling video-rate nanoscale imaging of a variety of biological phenomena that were previously inaccessible.[20-33]

The emergence of HS-AFM has dramatically increased the volume of acquired data. While conventional AFM experiments typically generate only a few tens of images, HS-AFM routinely produces thousands of frames in a single experiment. Consequently, HS-AFM requires analysis software with an architecture fundamentally different from that of conventional AFM.

Several studies have reported software tools designed to handle large-volume HS-AFM datasets by implementing functions such as automatic drift correction and image leveling.[34-36] In our previous studies, Kodec[37] was used for rapid inspection of acquired datasets, whereas quantitative analyses were performed using general-purpose image analysis software such as ImageJ[34,38] and Falcon Viewer,[39] custom software developed on the Igor platform.

However, in recent years, further improvements in HS-AFM instrumentation have enabled increasingly rapid data acquisition,[40-44] resulting in even larger datasets being generated over shorter time periods. Although existing tools provide essential functionality, these advances are making their processing capabilities and usability increasingly insufficient for current HS-AFM workflows. Thus, analysis software must evolve in parallel with hardware advances to provide faster image processing, more sophisticated graphical user interfaces (GUI), and improved usability.

To overcome these limitations, dedicated standalone software specialized for HS-AFM data analysis is highly desirable. Such software should provide efficient data-management functions that enable users to search and organize desired datasets from the large volume of AFM images. In addition, it should support an integrated workflow in which image correction, color-scale standardization, drift correction, quantitative analysis, and final export of publication-quality images and movies can be performed within a single environment.

AFM images often contain characteristic noise and distortions arising from scanner drift and feedback-response limitations, making image correction an essential preprocessing step.[13,35] In particular, color-scale standardization and drift correction are indispensable for generating interpretable movies and for comparing image sequences consistently. While manual correction of individual images may be feasible for conventional AFM datasets, the large number of images generated by HS-AFM makes such an approach impractical. Therefore, automated algorithms that minimize user intervention while maintaining flexibility for diverse datasets are strongly required.

Beyond image visualization, HS-AFM studies frequently require extraction of quantitative parameters such as distance, height, volume, and angle from image data, followed by analysis of their temporal variations.[20,26,32,45-51] Performing such analyses manually is labor-intensive and becomes increasingly impractical as dataset sizes grow. Efficient and integrated quantitative-analysis tools are therefore essential for fully exploiting HS-AFM datasets.

To address these challenges, we developed Ultrafast Microscopy Exploration (UMEX) Viewer, an integrated analysis platform for HS-AFM. The software incorporates knowledge and experience accumulated through years of HS-AFM research and has already been applied to a wide range of recent HS-AFM studies.[27-29,48,50-71] In this article, we describe the core analysis capabilities of UMEX Viewer and present practical workflows for commonly used HS-AFM data-analysis procedures.

## 2. Basic Functions and GUI

### 2.1. Software Components

UMEX consists of multiple applications designed for different types of HS-AFM data analysis. The software suite includes "UMEX Viewer" for general image analysis, image export, and movie generation; "UMEX Line Analyzer" for molecular shape analysis based on vector-line representations; and "UMEX Height Analyzer" for pixel-based height measurements. Together, these applications provide an integrated environment for efficient HS-AFM data analysis.

### 2.2. Supported File Formats

Since HS-AFM generates thousands of image frames during a single experiment, our HS-AFM systems use a proprietary ASD file format for AFM scan data, which stores multiple image frames acquired under the same measurement conditions in a single file. UMEX Viewer is primarily intended for the analysis of ASD datasets and also supports the ESD file format, which is used to store image-processed data exported from UMEX Viewer or Kodec. Because ESD files have the same internal structure as ASD files, they preserve quantitative height information without reducing the height gradation, unlike bitmap images. Multi-frame 32-bit floating-point TIFF images are also supported, allowing data exported from ImageJ to be loaded directly.

## 2.3. Software Architecture

UMEX Viewer is a standalone software package developed on the Microsoft .NET Framework 4.x platform. The .NET Framework was selected because it combines near-native execution performance with rapid development of sophisticated GUIs. These features are well suited to HS-AFM data analysis, which requires interactive visualization and efficient handling of large image datasets. Although platforms such as Igor Pro, ImageJ, LabVIEW, and MATLAB provide extensive libraries and analysis tools, a standalone .NET-based implementation offers greater flexibility for high-throughput image processing and specialized GUI development.

Another advantage of the .NET Framework is that its runtime is included by default in most Windows environments; therefore, no additional runtime installation or third-party software licenses are typically required. Furthermore, UMEX Viewer has no external library dependencies and can be used simply by copying the executable files. The only exception is movie export, which requires FFmpeg (see Section 5.2).

In the current development version, UMEX Viewer has been migrated from the legacy .NET Framework to the modern .NET platform (currently .NET 8), enabling compatibility with modern machine-learning ecosystems and providing a foundation for future AI-assisted image analysis. Although UMEX Viewer is primarily developed for Windows, it can also be operated on macOS through virtualization software or compatibility layers.

## 2.4. Graphical User Interface

Figure 1(a) shows the main interface of UMEX Viewer. A key feature of the software is its single-window interface, in contrast to the multi-window layouts commonly used in AFM analysis software. This design reduces window clutter during the analysis of multiple files and facilitates efficient data management.

In the upper-left area, images from channels 1 and 2, which are typically assigned to trace and retrace images, respectively, are displayed. Below these images are the "Seek Bar," which indicates the currently displayed frame and allows direct navigation to a selected frame, and the "ASD File Browser," which lists the loaded files. In the lower-left area, the "Image Processing Parameters" section contains functionally categorized tabs that provide access to various image-processing operations.

In the upper-right area, the "File Properties" section displays imaging parameters, including pixel size, scan size, frame time, and scan rate. Additional information can be viewed in "Extra Info," which is accessible from the right-click context menu of "File Properties" (Supplementary Fig. 1(a)). In the lower-right area, the "Frame Properties" section shows the scan area and the probe trajectory. Information stored in the ASD file, such as the acquisition time for each frame and amplitude values used to estimate tip–sample interaction forces,[72,73] is also displayed.

ASD files can be opened by dragging and dropping them anywhere within the main window. Alternatively, after registering the ASD extension using the file-association function, files can be opened directly by double-clicking them in the file explorer. When a single ASD file is opened, all ASD files in the same folder are automatically loaded and listed in the "ASD File Browser" (Fig. 1(b)). For each file, the browser displays the file name, acquisition time, number of frames, frame time, and scan size, allowing users to identify and select the desired dataset easily. A file can be opened immediately with a single click in the “ASD File Browser.”

The “ASD File Browser” also includes a reload function that adds newly acquired files to the list without requiring the software to be restarted. This enables data to be viewed and analyzed while measurements are still in progress. In addition, file-management operations, including file searching and renaming, are available from the right-click context menu (Supplementary Fig. 1(b)).

The currently displayed frame is indicated by the “Seek Bar,” and clicking the bar moves directly to the selected frame. Frames can also be navigated using the left and right arrow keys, while the up and down arrow keys switch between files. Faster frame navigation is available by holding down the Shift and/or Control key.

All numeric input boxes support mouse-wheel adjustment when the cursor is placed over them, and fine tuning can be performed by scrolling the wheel while holding the right mouse button. Unlike conventional analysis software, parameter changes are applied immediately to the displayed image without requiring an explicit update or execution step. This real-time feedback enables more efficient and intuitive data analysis. Various display parameters can be saved as parameter files and restored after restarting the software, enabling reconstruction of identical display conditions (Supplementary Fig. 1(c)).

## 2.5. Search and Filtering Functions

When working with large collections of HS-AFM datasets, an efficient search function is essential for locating files of interest. The search function can be accessed via the right-click context menu in the file list (Fig. 1(c)). It enables users to search for ASD files within folders containing a large number of datasets based on criteria such as number of frames, frame time, scan size, and pixel dimensions. Only files that satisfy the specified criteria are then loaded and displayed in the file list.

## 2.6. Data Loading and Display Time

For browsing large HS-AFM datasets, minimizing the time required to load and display each frame is one of the most important factors. On a standard personal computer, loading and displaying a single frame requires approximately 20 ms. Even when masked flattening is applied (Section 3.4), the processing time is only about 30 ms per frame, allowing image browsing at video rate. For images smaller than 300 × 300 pixels, the overall display time changes little because the processing time is dominated by image analysis rather than file loading.

## 2.7. Overview of ASD Data Analysis

Figure 1(d) summarizes the workflow of ASD data analysis. ASD files are first selected from the file list and loaded for analysis, after which basic image-processing operations can be applied. Subsequently, users can either export images and movies or perform more detailed image analysis. For image and movie export, operations such as color-scale adjustment, frame-range selection, and drift correction can be performed prior to saving in the desired format.

For quantitative analysis, basic functions are available through the “Basic Options” in the “Detail Analysis” section (Fig. 1(d)), while more specialized frame-by-frame analyses can be performed using the “Advanced Options” (UMEX Line Analyzer and Height Analyzer). Alternatively, data after image-processing operations, such as flattening and drift correction, can be exported in the ESD format or as multi-frame 32-bit floating-point TIFF images for analysis using external software.

# 3. Image Correction Functions

## 3.1. Image Filtering

UMEX Viewer implements a range of basic image filtering functions commonly used in AFM data analysis (see the "Image Filtering" section in Fig. 1(d)). Among these, flattening, also known as leveling, is one of the most essential preprocessing steps in HS-AFM analysis because AFM images often contain height artifacts. The details of flattening are described in the following section.

Depending on experimental conditions, HS-AFM images may contain significant noise that obscures structural details. Gaussian filtering is frequently used for noise reduction. Figure 2(a) shows noisy a HS-AFM image of ring-shaped Smc5/6 bound to DNA.[62] Gaussian smoothing effectively reduces noise (Fig. 2(b)), although excessive filtering may compromise spatial resolution.

Further noise reduction can be achieved by frame averaging, in which neighboring frames are averaged in the temporal domain.[53] This processing makes the ring-shaped molecular structures more clearly visible (Fig. 2(c)). In conventional AFM, frame averaging is less commonly used because the long acquisition time required for each image often necessitates frame-to-frame registration to correct for drift. In contrast, HS-AFM typically operates at frame rates above 1 fps, making such registration unnecessary in many cases. However, because frame averaging reduces temporal resolution, it is mainly suitable for enhancing the visibility of structural features in static images. Care should be taken when exporting movies, as frame averaging may produce pronounced motion-blur-like artifacts. In addition to the conventional rectangular kernel, a triangular weighted kernel is available, which assigns progressively lower weights to frames farther from the target frame.

Although not used in all analyses, UMEX Viewer also implements several specialized filtering and correction functions to address specific imaging conditions. Rolling-ball processing is useful for removing background variations and enhancing small features of larger surface structures such as viruses.[53] For visualization of molecules at different height levels or enhancement of backbone structures, Laplacian filtering and edge detection are effective.[74] FFT filtering suppresses noise and enhances periodic features (see Section 4.4). The software also implements tip deconvolution to correct for tip-induced broadening effects.[6,75-77] In addition, hysteresis correction is available and may be required for large scan ranges;[78] however, it becomes unnecessary when feedforward control is applied in the control software.[79]

## 3.2. Flattening: Overview

AFM images often contain apparent height artifacts arising from several factors described later. Although HS-AFM is considerably less susceptible to drift than conventional AFM because of its much higher frame rate, flattening remains essential for reliable visualization and quantitative interpretation.

Flattening consists of two main procedures: plane correction and line-by-line correction. Plane correction removes global tilt using a two-dimensional polynomial model. In contrast, line-by-line correction compensates for scan-line–dependent height variations along the X and/or Y directions.[13,34,35,77,80]

Representative cases are shown in Figures 2(d–f). First, when slope correction during data acquisition is incomplete, or when image tilt varies from frame to frame, plane correction can be used to flatten the images (Fig. 2(d)). In addition, unstable tip conditions may cause height offsets that vary between individual scan lines. In such cases, line-by-line flattening along the X direction (rows) effectively removes these artifacts (Fig. 2(e)). Furthermore, periodic vertical stripe artifacts may appear when the resonance of the X piezo scanner is excited during high-speed scanning. These artifacts can be corrected by line-by-line flattening along the Y direction (columns) (Fig. 2(f)).

Conventional AFM often permits manual adjustment of flattening parameters on an image-by-image basis, whereas HS-AFM demands robust algorithms applicable to large datasets with minimal parameter tuning. For this reason, in addition to conventional least-squares fitting, we implemented two robust estimators, Theil–Sen[81,82] and Random Sample Consensus (RANSAC).[80,83] Theil–Sen reduces sensitivity to outliers through median-based estimation. For zero-order correction, the median height is used as a robust estimate of the offset. RANSAC iteratively fits models using inliers within a specified threshold, enabling robust correction based on automatically identified near-flat substrate regions.

### 3.3. Flattening without Masking

Figure 3(a) shows a flowchart of the flattening procedure. In general, applying an exclusion mask and subsequently repeating the flattening process provides better results than using the initially flattened image alone.[35] Although iterative masking can further improve accuracy, it also increases computational cost. Therefore, the performance of the flattening algorithm under non-masked conditions remains important. To assess this, we first compare different fitting methods under non-masked conditions, focusing on three representative cases commonly encountered in HS-AFM imaging.

We first analyzed an isolated Smc5/6 molecule using XY line-by-line flattening (Fig. 3(b), Movie S1(a)). Least-squares fitting produces artificial dark lines in scan lines containing molecules, referred to here as baseline-offset artifacts. In contrast, the Theil–Sen estimator suppresses such artifacts because scan lines are typically dominated by background pixels, although weak residual artifacts remain. RANSAC further reduces baseline-offset artifacts but is more sensitive to noise and may introduce pronounced line noise not present in the original image. Overall, for sparsely distributed molecules with large exposed substrate areas, both Theil–Sen and RANSAC outperform least-squares fitting.

We next examined cases with high molecular coverage (Fig. 3(c), Movie S1(b)). Least-squares and Theil–Sen provide global correction but still produce baseline-offset artifacts. In contrast, RANSAC becomes unstable and produces pronounced line artifacts due to fragmented background regions. Even for isolated molecules, RANSAC can become unreliable in the presence of surface modifications or noisy substrates, indicating limited applicability of line-by-line RANSAC flattening.

In contrast, RANSAC can be advantageous for plane correction under specific conditions. As shown in Fig. 3(d), Movie S1(c), a lipid membrane surface containing two flat regions separated by a step structure was analyzed. Least-squares and Theil–Sen introduce residual tilt, whereas RANSAC correctly identifies the flat regions and achieves accurate flattening of both regions.

Overall, for non-masked flattening, the optimal algorithm depends on the image characteristics; however, the Theil–Sen estimator shows the most robust all-around performance across different conditions.

## 3.4. Flattening with Masking

This section describes flattening with masking, whose procedure consists of two steps (Fig. 3(a)). First, a pre-flattening operation is applied, followed by generation of a mask. Post-flattening is then performed after excluding the masked regions, enabling selective flattening of the substrate region.

Although various masking strategies have been proposed for AFM image processing,[35,84-86] we adopted the following algorithm to achieve more robust mask generation. Here, we describe the procedure using the actin filament shown in Supplementary Fig. 2(a) as an example.

As described in Section 3.3, the Theil–Sen estimator often provides the best results for standard flattening. However, when it is applied to the pre-flattening operation, scar-like artifacts that should be excluded from the flattening process may not be sufficiently detected during mask generation. In such cases, least-squares fitting can be more suitable than the Theil–Sen estimator. Consequently, in this example, XY line-by-line least-squares fitting was used for the pre-flattening step.

To exclude regions other than the flat substrate from the flattening process, UMEX Viewer normalizes the height values of all pixels in the image to a range of 0–100% and defines regions above or below a user-defined threshold as the mask region. In this example, regions above the threshold were masked to exclude molecular features. The generated mask region can be displayed in green, allowing the user to visually inspect the mask setting (Fig. S1(b,c)).

In many cases, a threshold of approximately 50% provides good results. However, if the threshold is too high, part of the molecule is excluded from the mask region (Supplementary Fig. 2(b)), whereas if the threshold is too low, the mask region extends into the substrate region (Supplementary Fig. 2(c)). Therefore, the threshold must be adjusted so that the mask region appropriately covers the molecular features.

Moreover, even when an appropriate threshold is obtained for a single frame, applying the same threshold to multiple frames can result in inappropriate mask regions because of frame-to-frame variations in the height distribution and background shape.

To address this limitation, UMEX Viewer incorporates several additional processing steps. First, molecular regions are identified using a height threshold (Supplementary Fig. 2(d)), followed by mask smoothing and dilation. Smoothing removes small noise-induced spurious mask regions (Supplementary Fig. 2(e)), while dilation expands the mask to better cover molecular boundaries (Supplementary Fig. 2(f)).

Next, we analyzed the same datasets shown in Fig. 3(b–d) using masked flattening. As shown in Fig. 3(e), masked flattening of an isolated molecule effectively suppresses the baseline artifacts along scan lines containing molecules observed in the non-masked case. This improvement is particularly pronounced for the Theil–Sen estimator, which remains stable even when molecular regions are slightly underestimated (data not shown), resulting in nearly complete removal of residual dark-line artifacts. In contrast, RANSAC still exhibits pronounced line noise even under masked conditions.

For high-coverage cases (Fig. 3(f)), masking substantially improves substrate flattening for least-squares fitting and Theil–Sen estimation, resulting in nearly complete correction, whereas RANSAC still exhibits line-noise artifacts even under masked conditions. Plane correction on stepped surfaces was also evaluated under masked conditions (Fig. 3(g)). In all cases, masking improves performance, and all three methods show better flattening results than non-masked processing.

In summary, the Theil–Sen estimator provides the most robust performance across diverse surface conditions, regardless of masking. However, line-noise artifacts similar to those observed with RANSAC may occasionally persist even after applying the Theil–Sen estimator. In such cases, least-squares fitting can sometimes yield better results and should also be considered.

### 3.5. Manual Flattening Masking

In some cases, it can be difficult to find a single set of parameters for automatic flatten-mask generation that fully covers molecular regions in all frames, resulting in incomplete flattening. To address this limitation, a manual masking mode ("Flatten Exclude Region") was also implemented. Supplementary Fig. 3(a) shows an AFM image of Smc5/6. In this mode, the mouse cursor is converted into a circular brush, allowing users to define mask regions by left-clicking (Supplementary Fig. 3(b)). The brush size can be adjusted using the mouse wheel (Supplementary Fig. 3(c)). Right-clicking erases masked regions covered by the brush (Supplementary Fig. 3(d)), while middle-clicking clears all masks (Supplementary Fig. 3(a)).

Mask regions are stored on a frame-by-frame basis, allowing them to be restored during frame navigation. The entire mask configuration can also be saved as a parameter file (see Fig. 1(d)), enabling recovery of the same analysis state after restarting the software.

### 3.6. Color Contrast

In AFM images, height information is displayed using false-color mapping; therefore, appropriate color settings are essential for clearly visualizing molecular structures. This software provides multiple colormaps, including the gold colormap commonly used in HS-AFM.

Conventional AFM software generally uses linear scaling based on the minimum and maximum intensity values in each image (referred to in UMEX Viewer as the “Norm Range” algorithm). Pixel values are normalized to this range, and colors are assigned according to user-defined relative values. Although Norm Range is intuitive for manual adjustment, manually optimizing the color scale for each frame is impractical for large HS-AFM datasets. Furthermore, the method is highly sensitive to outliers, and even small amounts of noise or imaging artifacts can significantly affect image contrast and cause frame-to-frame variations.

To address these limitations, the software introduces five additional scaling modes: Percentile, Peak Ref, Zero Ref, Max Ref, and Min Ref. These should be selected according to the image characteristics (Fig. 4(a)).

For routine image inspection, where a fixed color scale is not required, the Percentile mode generally provides good results. In this mode, all pixel values are sorted, and specified percentiles (default: 99.8% and 0.5%) are used as the upper and lower bounds of the color scale. Compared with Norm Range, this approach reduces the influence of outliers and provides more stable contrast.

However, Percentile scaling still causes frame-to-frame color variations. As shown in Fig. 4(b), actin filaments are clearly visible in the first frame, whereas their contrast is noticeably reduced in the second frame (Movie S2(a)). This occurs because the appearance of scars and debris shifts the maximum intensity value, resulting in an expansion of the color range from 11 nm to 26 nm. Such automatic changes in the color scale are undesirable when exporting images or movies, where consistent visualization across frames is required. Therefore, methods for fixing the color scale are necessary.

Figure 4(c) shows the height-distribution profiles of the two images shown in Fig. 4(b). In both profiles, a prominent substrate-derived peak is observed because a large fraction of the image consists of exposed substrate regions. The Peak Ref algorithm fixes the color scale using this substrate peak as a reference (Fig. 4(d)). First, the substrate peak is detected, and a user-defined offset is applied to determine the minimum value. The maximum value is then calculated by adding a fixed scale width (Fig. 4(e)). This approach enables stable color scaling even in the presence of noise, debris, or structural changes (Movie S2(b)).

Peak Ref works well when the substrate region is widely exposed. However, when the surface is nearly fully covered by molecules, as shown in Fig. 4(f), Movie S2(c), the height distribution can change significantly between frames, as seen in the second image. As indicated by the histograms in Fig. 4(g), this instability arises because the dominant peak shifts between substrate-derived and molecule-derived populations due to changes in their relative coverage.

In such cases, the Zero Ref method is effective (Fig. 4(h), Movie S2(d)). This approach uses the background-corrected image obtained via the Flatten operation (see Section 3.2–3.5), and defines the resulting zero level as the reference height (Fig. 4(i)). When Flatten is disabled, the median height of the entire image is used as the reference.

Another limitation of Peak Ref occurs in cases where flat substrate regions and step structures coexist, as shown in Fig. 4(j), Movie S2(e). If one of the substrate regions remains sufficiently dominant in all frames, Peak Ref can still perform well. However, when the dominant peak alternates between frames (Fig. 4(k)), the reference height changes accordingly, resulting in frame-dependent color variations.

In such cases, the Peak Select mode can be used to maintain a fixed color scale by selecting either the higher or lower peak as the reference (Fig. 4(l), Movie S2(f)). In the present example, the lower peak is used as the reference (Fig. 4(m)). Alternatively, the Zero Ref mode can be used when the same surface region can serve as the zero-height reference across all frames after masked flattening.

In addition, the software implements Max Ref and Min Ref, which define the reference height based on the upper or lower percentile limits of the height distribution. When the above methods do not provide satisfactory correction, these approaches may also be useful.

# 4. Data Analysis Functions

## 4.1. Line Profile

Line-profile analysis is one of the most commonly used methods for analyzing HS-AFM data. This mode extracts height-profile data along a user-defined line specified by control points (Fig. 5(a1)) and displays the resulting profile on a chart (Fig. 5(a2)). This allows users to quantitatively estimate molecular heights, apparent widths, interdomain distances, and other structural parameters.[30-32,52,55,57,68,87] In UMEX Viewer, when the frame is changed, the profile is automatically updated according to the newly displayed image, eliminating the need to redefine the line for each frame and enabling efficient analysis. Line profiles can be exported as bitmap images, ASCII data, or vector graphics. The vector format allows direct insertion into Illustrator or PowerPoint without loss of resolution, while fonts and line widths remain editable in Illustrator.

To facilitate accurate measurements, vertical and horizontal marker functions were implemented to display height differences and lateral distances on the profile plot (Fig. 5(a2)). The region defined by the horizontal markers is also displayed on the AFM image. In general, molecular lateral size can be measured more reliably by drawing a line profile extending slightly beyond the target structure width and positioning the horizontal markers at the full width at half maximum (FWHM) of the target structure in the height profile, rather than placing the line endpoints at the apparent molecular edges and directly reading the line length.

For filamentous structures such as DNA and cytoskeletal filaments, line-profile analysis along the molecular contour is often required to measure contour length, analyze periodic pitch structures, or determine the positions of bound proteins. Figure 5(b) shows an example of line-profile analysis performed on a curved actin filament. In UMEX Viewer, filament contours can be defined using a polyline consisting of multiple control points (see arrows in Fig. 5(b1)). Unlike conventional software, users can select an existing control point and press the Enter or Delete key to insert or remove a point, respectively.

When only a few control points are used, the line may deviate from the filament and extend into the substrate region (Fig. 5(b1)). Consequently, accurately tracing the molecular contour often requires many control points, which becomes laborious when repeated across multiple frames (see Section 6.3 for an example analysis). To address this issue, the software provides a spline-interpolation function that automatically smooths the polyline into a curved contour, allowing accurate tracing of filamentous structures with only a few control points (Fig. 5(b2)). In addition, line-width averaging (Fig. 5(b3)) and profile-smoothing functions are available to improve the quantitative analysis of filament pitch structures (Fig. 5(b4)). For helical filaments, the line-width averaging direction can be sheared so that averaging is performed along the helical pitch direction rather than strictly perpendicular to the filament axis.

## 4.2. Height-Distribution Analysis

Statistical analysis of the height distribution within an AFM image is widely used in both conventional AFM and HS-AFM data analysis. In UMEX Viewer, this function is referred to as "Statistics." Figure 5(c) shows an example of its application to an AFM image of a GroEL monolayer formed on mica. The height distribution can be displayed as a conventional histogram or as a probability density profile generated using kernel density estimation (KDE). As with the other analysis functions, both the histogram and KDE profile are automatically updated when the displayed frame is changed.

This analysis provides several types of quantitative information. For example, surface roughness can be estimated from the standard deviation of the height distribution, whereas differences between distinct peaks can be used to determine molecular heights or surface step heights. Height histograms are therefore also frequently used for Z-piezo calibration. In UMEX Viewer, the mean height ($\mu$), standard deviation ($\sigma$), skewness (S), and kurtosis (K) are automatically calculated and displayed, facilitating quantitative characterization of the distribution and assessment of its deviation from a normal distribution.

In HS-AFM data analysis, histogram analysis is particularly useful for evaluating image-contrast settings when the automatic color-scaling methods described in Section 3.6 do not produce satisfactory results (for example, the right panels of Figs. 4(f) and 4(j)). Furthermore, it is useful for optimizing flattening procedures, as properly flattened images typically exhibit a narrower and sharper height-distribution peak.[35]

## 4.3. Fourier Analysis

AFM topography images contain frequency components arising not only from surface topography but also from mechanical vibrations, electrical noise, and feedback oscillations. This function evaluates the frequency spectrum of an AFM image by performing fast Fourier transform (FFT) analysis on a time-series signal generated from concatenated scan lines. The resulting spectrum enables identification of periodic noise components and is therefore useful for diagnosing instrumental problems.

It can also be used to estimate the feedback bandwidth of an AFM system,[17,41,43] which determines the maximum achievable frame rate and minimum detectable force.[73] Although the feedback bandwidth can be measured by applying a modulation signal to the Z piezo under engaged conditions and performing a frequency sweep,[17] this approach requires additional hardware and a separate measurement from imaging. In contrast, the noise spectrum provides a simpler estimate because it can be obtained during routine imaging, as described below.

Figure 5(d) shows an AFM image of a lipid membrane domain under stable feedback conditions. The image was acquired at 100 μm/s over a 100 × 100 nm² area with 800 × 100 pixels. As described later, these imaging conditions resulted in vertically elongated non-square pixels, rather than the conventional square pixels typically used in AFM imaging. The frequency spectrum obtained from this image is shown below the AFM image. The red curve was obtained with a smoothing coefficient of 10 and shows multiple peaks at the X-scan frequency of 0.5 kHz and its harmonics. To suppress these scan-related peaks, the smoothing coefficient was increased to 80, producing an envelope spectrum that gradually decays with increasing frequency (blue curve). The cutoff frequency can be identified from the onset of this decay. The feedback bandwidth can then be estimated as approximately one-half to one-third of the cutoff frequency.[73] In this example, the cutoff frequency is approximately 55 kHz, corresponding to an estimated feedback bandwidth of approximately 20 kHz.

Although this method is useful for analyzing existing AFM datasets, it has several limitations. First, the apparent bandwidth depends on the feedback parameters and may vary with imaging conditions. Second, when the spectral intensity decreases gradually, determination of the cutoff frequency becomes ambiguous, making accurate bandwidth estimation difficult.

For direct experimental measurement of the bandwidth, the feedback-oscillation frequency can provide an alternative estimate. As the feedback gain is increased, the AFM feedback system typically becomes unstable and oscillates near its bandwidth limit. Figure 5(e), top, shows an AFM image acquired from the same region as in Fig. 5(d) after intentionally increasing the feedback gain to induce feedback oscillation. Fourier analysis of this image, shown below the AFM image, reveals a broad peak corresponding to the oscillation frequency, in addition to peaks originating from the scan frequency (Fig. 5(e), bottom). Because this peak is well defined, its frequency can be determined accurately as 55 kHz, the same as the cutoff frequency identified in Fig. 5(d).

It should be noted that these analyses require the following condition to be satisfied:

$$f_{\mathrm{Nyquist}} \gg f_{\mathrm{c}}, \tag{1}$$

where $f_{\mathrm{c}}$ is the cutoff frequency of the feedback bandwidth. $f_{\mathrm{Nyquist}}$ is the Nyquist frequency, corresponding to the maximum frequency in the spectrum. If this condition is not satisfied, high-frequency components are aliased into the lower-frequency region, preventing the spectral intensity from decaying at high frequencies and making determination of $f_{\mathrm{c}}$ impossible.

$f_{\mathrm{Nyquist}}$ is given by

$$f_{\mathrm{Nyquist}} = \frac{f_{\mathrm{sample}}}{2}, \tag{2}$$

where $f_{\mathrm{sample}}$ is the sampling frequency, calculated as

$$f_{\mathrm{sample}} = \frac{v_{\mathrm{scan}} N_x}{W_{\mathrm{scan}}}. \tag{3}$$

Here, $v_{\text{scan}}$, $W_{\text{scan}}$, and $N_x$ are the scan velocity, scan size in the X direction, and number of pixels in the X direction, respectively. $f_{\text{sample}}$ used for image acquisition can also be checked in "Extra Info," which can be opened from the File Properties context menu in UMEX Viewer (Fig. 1(a)).

In practice, $v_{\text{scan}}$ is often kept constant to maintain stable feedback conditions, regardless of $W_{\text{scan}}$. Therefore, when $N_x$ is fixed, $f_{\text{sample}}$ decreases as $W_{\text{scan}}$ increases. Accordingly, to satisfy Eq. (1), $W_{\text{scan}}$ must be sufficiently small. For example, in the case shown in Fig. 5(d,e), $f_c$ is 55 kHz, and $f_{\text{Nyquist}}$ should therefore be approximately 150 kHz or higher. Under typical HS-AFM imaging conditions, such as $v_{\text{scan}}$ = 100 μm/s and $N_x$ = 100 pixels, $W_{\text{scan}}$ must be kept below 30 nm.

However, when the scan size is too small, it often becomes difficult to judge whether appropriate feedback conditions are being maintained. Therefore, in the experiment shown in Fig. 5(d,e), only $N_x$ was increased to 800 while the number of pixels in the Y direction was kept at 100. This resulted in non-square pixels rather than conventional square pixels. This approach allows $f_{\text{sample}}$ to be increased without reducing either $W_{\text{scan}}$ or the frame rate.

This approach is also practical for routine imaging when combined with Gaussian filtering during subsequent image processing, which effectively suppresses aliased high-frequency noise and improves image quality. In UMEX Viewer, the image aspect ratio is determined from the scan size, allowing data acquired with non-square pixels to be displayed correctly. However, care should be taken when using external software, as some programs may display such images with an incorrect aspect ratio.

### 4.4. Two-Dimensional FFT Analysis and Filtering

The two-dimensional (2D) FFT analysis mode enables real-time visualization of the 2D Fourier spectrum of each AFM frame. Figure 5(f) shows a representative analysis using a 2D Annexin V crystal. Although the original AFM image is partially obscured by measurement noise, the crystal lattice can still be recognized (Fig. 5(f1), Movie S3(a)). Applying a 2D FFT transforms the image into Fourier space, where characteristic diffraction-like spots corresponding to the crystal periodicity become visible (Fig. 5(f2)). This allows periodic structures to be identified and their spatial frequencies to be analyzed.

The FFT filtering function extracts periodic components by masking Fourier components whose intensities fall below a user-defined threshold (Fig. 5(f3)). Applying an inverse FFT to the filtered spectrum reconstructs an image in which periodic structures are emphasized by suppressing non-periodic components (Fig. 5(f4), Movie S3(b)).[79,88] Compared with conventional approaches that require manual selection of mask regions in Fourier space, this method provides automated processing while preserving non-periodic structural features. For example, the pit structures in the honeycomb lattice indicated by arrows in Fig. 5(f1) remain visible after FFT filtering (Fig. 5(f4)).

In HS-AFM data analysis, this functionality is particularly useful for piezo-scanner calibration using 2D crystals, as Fourier filtering enables reliable calibration even when the crystal lattice is poorly resolved because of a blunt AFM tip.[79]

## 4.5. Zoom Mode

HS-AFM enables high-resolution imaging over relatively large scan areas within a short acquisition time.[78] As a result, thermal drift is considerably reduced compared with conventional AFM, allowing large-area images to be acquired without the need for image stitching. The Zoom mode facilitates analysis of such wide-field, high-resolution images by allowing users to magnify any region of interest defined by a bounding box within an AFM image. The position and size of the bounding box can also be adjusted interactively using the mouse.

Figure 5(g), left, shows an AFM image of an Annexin V 2D crystal membrane acquired over an area of 2.45 × 2.45 μm² (1100 × 1100 pixels) (Movie S4(a)). The frame acquisition time was 35 s. Because of the large scan area, the crystal lattice is difficult to distinguish clearly in the full-field image. In contrast, the zoomed views shown in Fig. 5(g), right, and Movie S4(b) reveal well-resolved lattice structures in both selected regions. This functionality is particularly useful for analyzing local structural details within large-scale images, as well as for examining individual molecules while retaining information about the overall sample morphology.[78]

## 5. Data Export and Conversion

### 5.1. Drift Correction by Cross-Correlation

During AFM imaging, the apparent position of a molecule of interest may change over time because of XY drift of the microscope or molecular diffusion.[16,34,36] As a result, direct visualization of raw image sequences often makes it difficult to distinguish genuine structural changes from translational motion. As a preprocessing step for image and movie export, this mode compensates for such motion by tracking a reference structure using cross-correlation.

Figures 6(a–c) and Movie S5 illustrate the drift-correction workflow using HS-AFM data of two Smc5/6 complexes bound to DNA.[62] First, the user selects a reference structure using a bounding box (Fig. 6(a)). Any structure that exhibits minimal positional and morphological changes during the observation period can be used as a reference, including either the target molecule itself or an unrelated object such as debris. In this example, a stationary background feature, namely an unruptured lipid vesicle, was selected as the reference structure. The software then uses the reference structure in the first frame as a template and performs cross-correlation-based image registration for all subsequent frames. Based on the calculated displacement, each frame is translated in the XY direction so that the reference structure remains centered throughout the image sequence (Fig. 6(b)). The image size and display position can subsequently be adjusted to generate a magnified view in which the target molecule remains centered in the field of view (Fig. 6(c)).

However, directly applying the drift trajectories obtained by cross-correlation often results in unstable correction, because the trajectories may be affected by image noise. Therefore, to improve robustness, Gaussian smoothing and median filtering can be applied to the calculated drift trajectories. The original and filtered trajectories can be compared directly within the time-chart display (Figs. 6(d,e)).

If the filtered trajectories are still insufficient for accurate correction, manual correction can be performed in the editing mode. Drift values for individual frames are displayed as control points and can be adjusted interactively by dragging with the mouse (Fig. 6(f)). Multiple frames can also be selected and modified simultaneously while holding the Shift key. These features allow flexible refinement of drift corrections for a wide range of HS-AFM datasets.

## 5.2. Image and Movie Export

The software can generate publication-quality images and movies directly from imported ASD files (Fig. 6(g)). Time stamps, scale bars, and height color bars, which are essential for AFM data presentation, can be automatically overlaid on the output images. Each element can be individually enabled or disabled, and its position and appearance can be customized by the user.

For image export, the software supports standard formats including PNG, TIFF, and BMP. Movie export is performed using FFmpeg[89] and supports multiple video formats, including WMV, MP4, MOV, and AVI (MJPEG, BMP). Because video encoding relies on an external FFmpeg executable, users must download FFmpeg separately and place the executable in the config folder before exporting movies.

## 5.3. 3D Rendering

The software can display AFM topographic images as interactive 3D surface models. As an example, Fig. 6(h) and Movie S6 show HS-AFM observations of the disintegration and regeneration of the upper ring of the MyD88 Toll/interleukin-1 receptor domain.[33] This function is implemented using OpenTK, a .NET wrapper for OpenGL. Similar to the 2D image display, the 3D model is automatically updated when the frame is changed, allowing time-dependent structural changes to be visualized intuitively. Users can rotate, translate, and inspect the 3D model using mouse operations, and the display can be further optimized by adjusting the height scaling and lighting parameters. Timestamps and scale bars can also be overlaid on the 3D view, and rendered 3D views can also be exported as image or movie files. This functionality facilitates intuitive interpretation of molecular structures and provides visually appealing representations for presentations and publications.[17]

## 5.4. File Conversion and Integration with Other Software

The software provides functions for converting ASD files into a variety of AFM file formats, including STP, ASC, BCR, GSF, and GWY. These formats can be imported into a wide range of analysis software packages, such as ImageJ, Igor Pro, Mathematica, WSxM, Gwyddion, SPIP, and Python-based analysis environments. The software also supports the import of files in these formats by converting them to ASD, allowing data saved using other AFM systems or software packages to be analyzed. Although ImageJ cannot read ASD files natively, direct import is possible through custom scripts.[34]

Because most AFM file formats store only a single frame per file, conversion of HS-AFM datasets can result in a large number of output files. This issue can be avoided by using the multi-frame 32-bit floating-point TIFF format, which stores all frames in a single file and can be directly imported into ImageJ.

# 6. Advanced Analysis

## 6.1. Overview

UMEX Viewer is suitable for image visualization and basic analysis; however, it does not support detailed structural quantification and statistical analysis of frame-by-frame measurements. To address these limitations, UMEX Height Analyzer and Line Analyzer were developed to quantitatively characterize molecular structures and their temporal changes. Height Analyzer treats AFM images as raster data and performs pixel-based analysis, whereas Line Analyzer represents structures as vectorized line objects and extracts geometrical and dynamic parameters. Table 1 summarizes the principal targets and quantitative information provided by these software packages.

**Table 1.** Analysis targets and quantitative parameters in UMEX Height Analyzer and Line Analyzer.

| | UMEX Height Analyzer | UMEX Line Analyzer |
|---|---|---|
| Data representation for analysis | Raster data | Vector data |
| Primary targets | Globular or arbitrarily shaped structures (e.g., protein complexes, aggregates) | • Geometrical features of linear molecules (e.g., DNA)<br>• Inter-domain distance analysis |
| Quantifiable parameters | • Position (x–y coordinates)<br>• Size (height, width, area, volume, perimeter)<br>• Circularity<br>• Diffusion coefficient | • Length (end-to-end distance, contour length)<br>• Shape descriptors (curvature, bending angle, radius of gyration)<br>• Filament pitch periodicity<br>• Kymographs |
| | | • Position and diffusion of DNA-binding proteins |
| Typical applications | Tracking and characterization of globular molecules | Structural and dynamic analysis of filamentous molecules |

### 6.2. UMEX Height Analyzer

This section describes UMEX Height Analyzer, which quantifies a variety of geometrical parameters from AFM images, including molecular height, area, volume, and other features summarized in Table 1. It has been used to analyze ligand binding and dissociation in biomolecules, as well as time-dependent structural changes in nucleic acid condensates.[50,65,70,71]

Figure 7(a) shows the GUI of the software. The AFM image is displayed in the upper-left panel, analysis parameters are configured in the lower-left panel, and quantified results are displayed in a table on the right. Right-clicking within the results table opens a context menu (Fig. 7(b)), which provides data-management functions, including deletion of analysis results for selected frames or all frames.

The software provides two analysis modes: Single-Particle Analysis, which analyzes a user-selected structure, and Multi-Particle Analysis, which automatically detects and quantifies multiple particles (Fig. 7(c,d)). The Multi-Particle Analysis mode also includes an auto-tracking function that enables simultaneous tracking and dynamic analysis of multiple particles (Fig. 7(e)).

In both modes, regions whose heights exceed a user-defined threshold relative to the substrate are recognized as particles. In the Multi-Particle Analysis mode, the substrate height is automatically estimated from the peak of the height histogram, similarly to the Peak Ref color-scaling method described in Section 3.4.

Figure 7(f) shows an example of Single-Particle Analysis applied to the Smc5/6 complex.[62] The user defines an analysis region, within which the detected molecular region is highlighted in purple (Fig. 7(f), middle). The analysis region is automatically tracked between frames but can also be adjusted manually when necessary (Movie S7). The user may additionally specify the substrate region manually (Fig. 7(f), right), which is useful for analyzing molecules adsorbed on scaffold structures.[65,74] Although only one particle can be analyzed at a time, this mode is particularly useful for analyzing specific domains within multi-domain proteins or rapidly moving molecules. Quantified parameters can be displayed as time charts and exported for further statistical analysis (Fig. 7(c)).

Figure 7(g) shows an example of Multi-Particle Analysis applied to lipid membrane domains. All domains in each frame are detected automatically and displayed using different colors. The measured value for each particle is displayed adjacent to the corresponding particle in the image. Statistical analysis can subsequently be performed using histograms of the measured parameters (Fig. 7(d)). This mode is mainly used for static structural analysis.

For analysis of dynamic structural changes, the auto-tracking function can be enabled. An example of this function is shown in Fig. 7(h), where streptavidin molecules adsorbed on a biotin-conjugated lipid membrane are independently detected and tracked over time. Particle trajectories are displayed as color-coded overlays, and particle correspondence between successive frames is determined using nearest-neighbor matching (Movie S8). Particle positions are also recorded, enabling dynamic analyses such as velocity or diffusion-coefficient measurements (Fig. 7(e)).[20,22,36,74]

## 6.3. UMEX Line Analyzer

This section describes UMEX Line Analyzer, which was previously referred to as "UMEX Viewer - Drift Analysis" in earlier publications.[48,57,59] Many biological molecules, including DNA and actin filaments, exhibit filamentous structures whose geometry and dynamics are important targets for quantitative analysis. In particular, numerous proteins bind specifically to filamentous substrates, such as transcription factors on DNA and cofilin on actin filaments.[37,90,91] Quantitative characterization of binding positions and associated structural changes is therefore important for understanding their biological functions.[90,91]

UMEX Viewer allows users to manually define line objects on AFM images and extract a variety of geometrical and dynamical parameters summarized in Table 1. Previous applications include analyses of inter-domain distances, molecular bending angles, DNA radius of gyration, kymographs of hairpin-DNA folding, and pitch periodicities of actin filaments.[48,51,57,59,62]

Figure 8(a) shows the GUI of the software. The AFM image is displayed in the upper-left panel, analysis parameters are configured in the lower-left panel, and quantified results are displayed on the right. As in UMEX Height Analyzer, the analysis results can be edited via a context menu accessible from the results table (Fig. 7(b)).

The software provides five analysis modes (Type A–E). In Type A–C analyses, quantitative parameters are extracted from the geometry of user-defined vector data (Figs. 8(b,c)), whereas Type D and E analyses are based on height information sampled along the defined lines. Both linear and circular line objects are supported, allowing analysis of circular molecules such as plasmid DNA (Fig. 8(d)).

Type-A analysis quantifies geometrical parameters of the entire line object, including contour length, inter-domain distance, bending angle, and radius of gyration (Fig. 8(e)).[10,21,26,31,45-47,49,62,91]

Multiple line objects can be analyzed simultaneously, and the results can be displayed in real time as time charts or histograms (Fig. 8(b)). Figure 8(f) shows an example of Type-A analysis applied to the Smc5/6 complex,[62] where distances between domains and arm-bending angles are quantified.

Type-B analysis evaluates local structural properties along the contour, such as curvature as a function of contour position (Fig. 8(g)). The results can be displayed as contour-position profiles or averaged over multiple frames (Fig. 8(c)). Similar contour-based analyses have previously been applied to the characterization of DNA curvature distributions and curvature changes induced by DNA-binding proteins.[10,45,49,91]

Type-C analysis extends Type-A and Type-B by allowing the position of a molecule bound to the filament to be specified (Fig. 8(h)).[90,91] This mode enables quantification of molecular dynamics, including position and diffusion coefficients, as a function of contour position or local curvature. Figure 8(i) shows an example applied to Smc5/6 molecules bound to DNA. This mode is particularly useful for analyzing proteins that undergo one-dimensional diffusion along filamentous substrates or repeatedly bind to and dissociate from such substrates.

Type-D analysis generates kymographs by extracting height profiles along a defined line from each frame and assembling them into a single image (Fig. 8(j)).[30,32,34,36,66,87,92] This enables visualization of time-dependent structural changes, such as those induced by protein binding. Because absolute height values may vary owing to imaging drift, flattening should be applied beforehand as described in Section 3.2–3.5.

Type-E analysis quantifies temporal changes in the helical pitch of filamentous structures such as actin filaments (Fig. 8(k)). Major and minor pitch periodicities are automatically detected from local maxima using user-defined pitch thresholds. Because these pitch structures are altered by actin-binding proteins, this mode enables quantitative analysis of their structural dynamics.[59]

For all analysis modes, line annotations are stored on a frame-by-frame basis and can be restored simply by revisiting a frame. All line positions can also be exported as parameter files and reloaded after restarting the software using functionality equivalent to that shown in Fig. 1(d), allowing the analysis state to be reproduced easily.

In principle, line generation could be automated through particle segmentation, as implemented in UMEX Height Analyzer, followed by skeletonization approaches. Although such methods are effective for relatively static structures in which individual objects are well separated,[10,13,90,91] reliable automatic analysis of HS-AFM data remains challenging because molecular motion may temporarily obscure parts of a structure or cause a single molecule to appear as multiple disconnected objects. In addition, this automated procedure assumes that lines are defined along molecular regions that are higher than the substrate. Therefore, some analyses, such as those shown in Fig. 8(f), left are difficult to automate because the lines are not defined along such molecular regions. For these reasons, manual annotation currently provides the most reliable results, although future advances in neural-network-based image analysis may enable further automation.

## 7. Conclusions

In this paper, we have presented UMEX Viewer, a software suite for visualization, processing, and quantitative analysis of HS-AFM data. The software was developed to address the increasing volume and complexity of modern HS-AFM datasets and provides an integrated framework for efficient data analysis. As HS-AFM technology continues to evolve, advances in instrumentation must be accompanied by corresponding developments in analysis software. Although UMEX Viewer was designed for HS-AFM, its underlying architecture is readily applicable to a wide range of AFM, particularly bioimaging analyses. We anticipate that the software will facilitate more efficient and quantitative analysis of dynamic biological processes and contribute to a deeper understanding of biological functions.

## Acknowledgements

We thank Professor Toshio Ando and Dr. Kien Xuan Ngo of Kanazawa University, Professor Takayuki Uchihashi of Nagoya University, and Dr. Takashi Morii of RIBM Co., Ltd., for their valuable support. We also thank Professor Hidehito Tochio of Kyoto University for providing the MyD88 Toll/interleukin-1 receptor domain sample. We also thank Ms. Aimi Makino, Ms. Kayo Nakatani, and Ms. Risa Omura for their technical assistance. We are grateful to numerous collaborators and users whose valuable feedback and suggestions have contributed to the continuous improvement of the software.

## Funding

This work was supported by the World Premier International Research Center Initiative (WPI), Ministry of Education, Culture, Sports, Science and Technology (MEXT), Japan; PRESTO, Japan Science and Technology Agency (JST) [JPMJPR20E3 and JPMJPR23J2 to K.U.]; KAKENHI, Japan Society for the Promotion of Science (JSPS) [25K09575 to K.U., 24H00402 to N.K., and 24K21942 and 25H00972 to M.S.]; and ERATO, JST [JPMJER2403 to M.S.].

## Competing interests

K. U. and N. K. serve as a technical advisor to Research Institute of Biomolecule Metrology Co. Ltd., a company that manufactures HS-AFM systems. The software described in this manuscript is distributed free of charge and is not limited to instruments from this company. The other authors declare no competing interests.

## Data availability statement

The datasets generated and/or analyzed during the current study are available from the corresponding author on reasonable request.

## Software availability

UMEX Viewer executable binaries and documentation are available at https://github.com/HS-AFM-developer.

# Figures

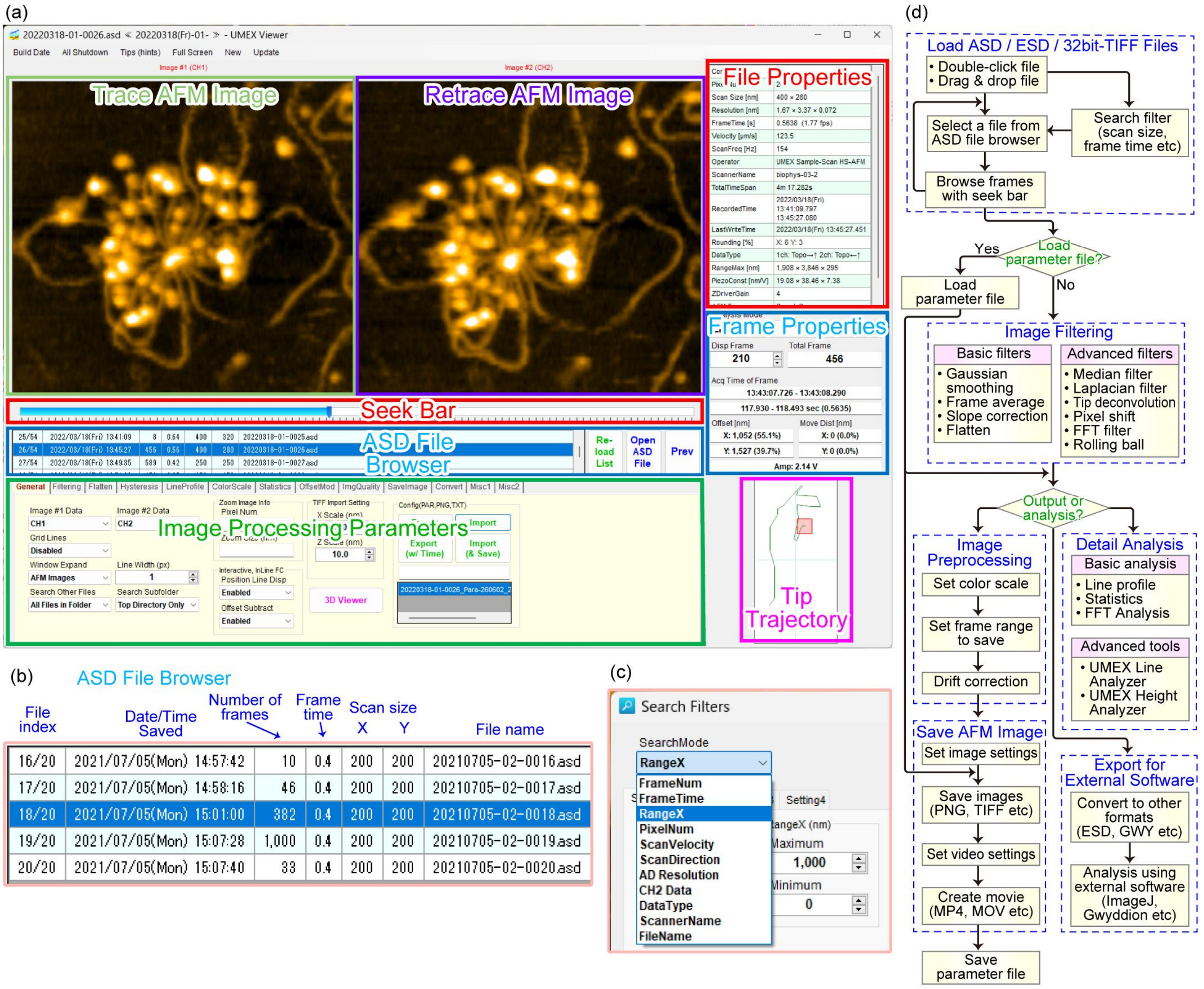

**Figure 1.** (**a**) Overall GUI of UMEX Viewer. (**b,c**) Enlarged views of the ASD file browser (b) and search filter function (c). (**d**) Flowchart of ASD data analysis, including data loading, image processing, data export, and quantitative analysis.

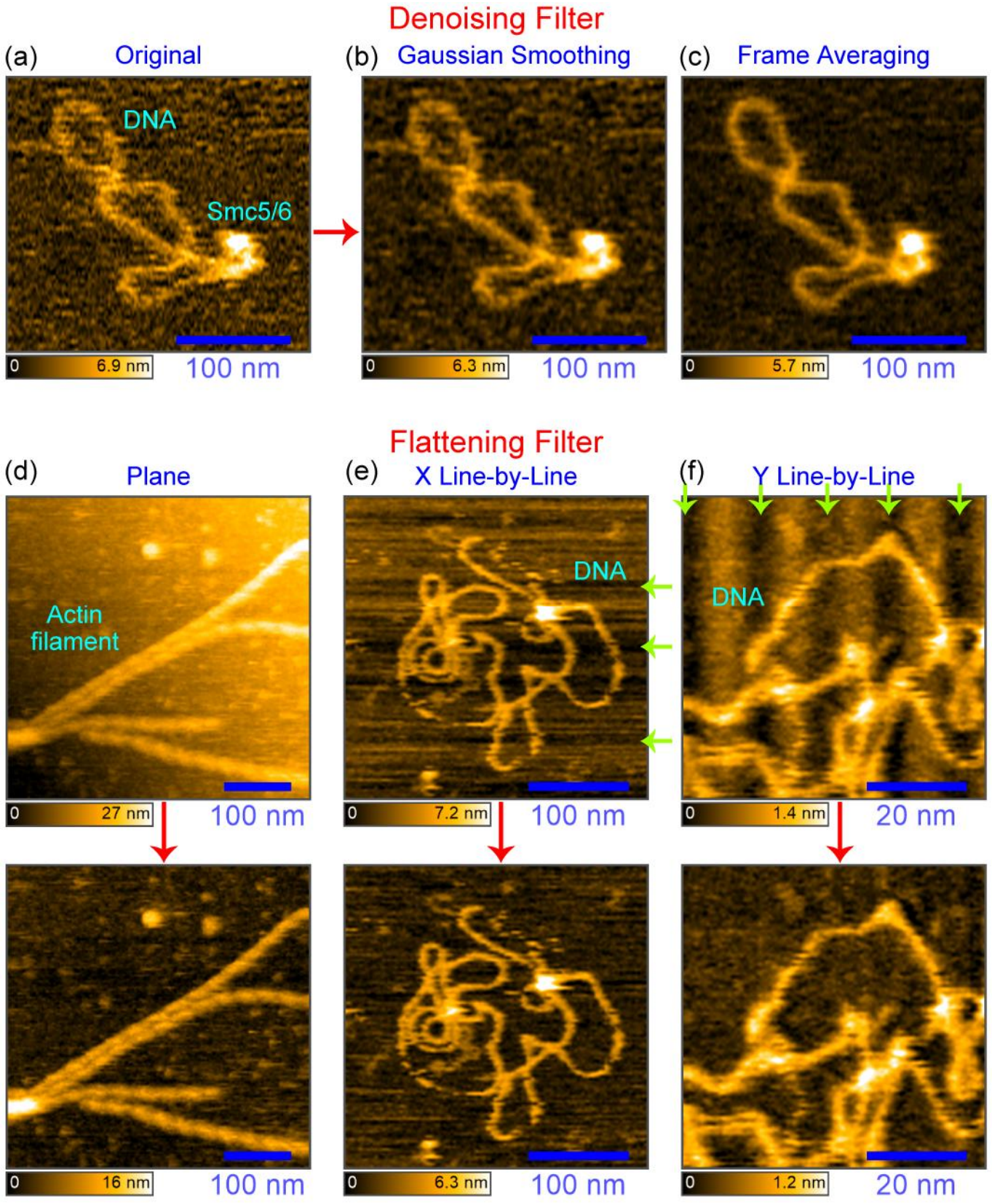

**Figure 2.** (**a–c**) HS-AFM images of Smc5/6 bound to DNA illustrating the effects of denoising filters: (a) original HS-AFM image, (b) Gaussian-smoothed image, and (c) frame-averaged image with a triangular kernel. (**d–f**) HS-AFM images showing the effects of plane flattening on actin filaments (d), X line-by-line flattening on DNA (e), and Y line-by-line flattening on DNA (f). In each panel, the upper image shows the original AFM image, and the lower image shows the flattened image.

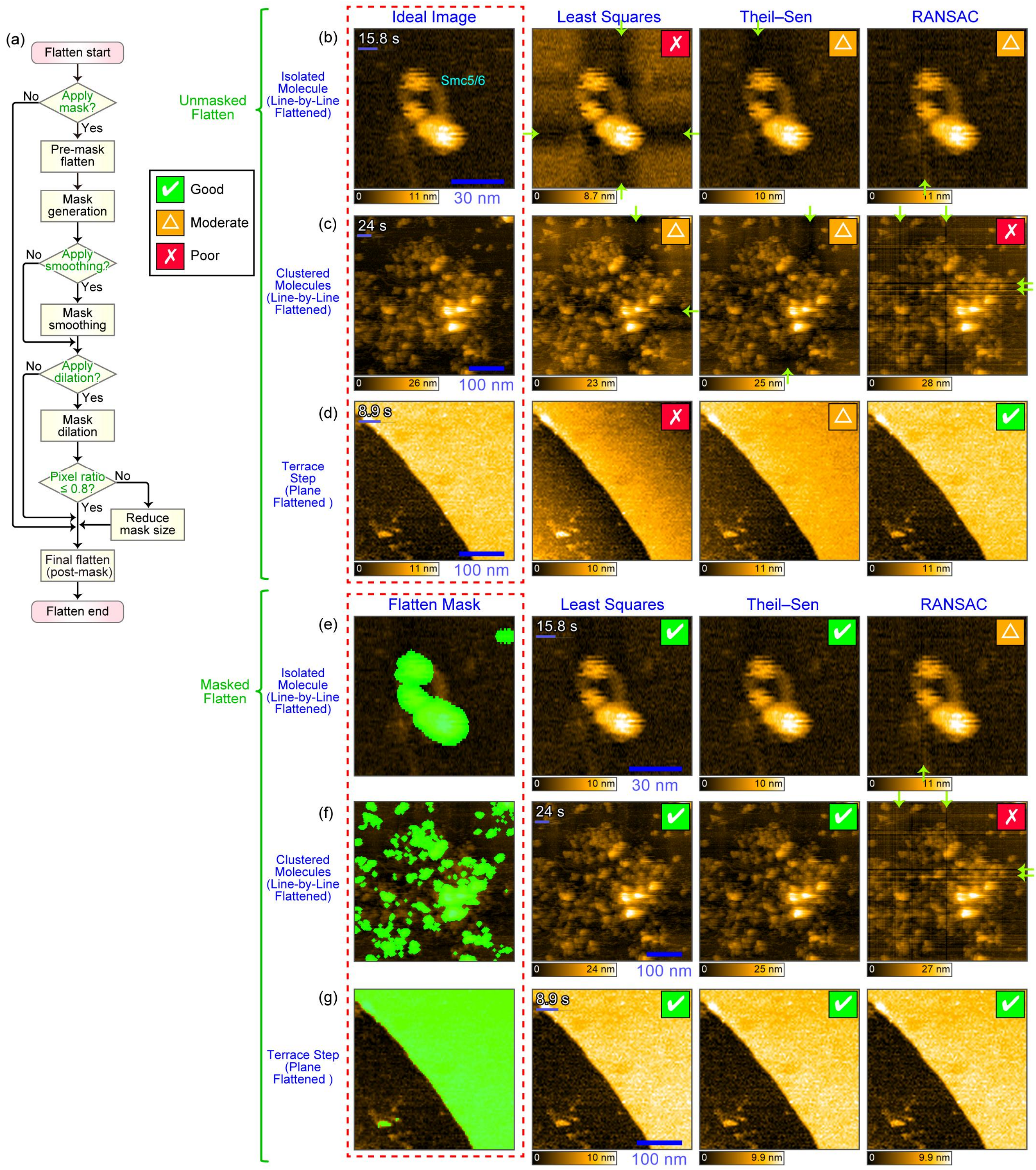

**Figure 3.** (**a**) Flowchart of the flattening procedure. (**b–d**) HS-AFM images of flattening without masking, including the application of XY line-by-line flattening to an isolated Smc5/6 complex (b) and clustered molecules (c), and plane flattening to a terrace step of a lipid membrane (d). From left to right, the panels show the ideally flattened AFM image and the flattening results obtained using

the least-squares, Theil–Sen, and RANSAC algorithms. (**e–g**) HS-AFM images of flattening with masking. The descriptions are the same as those in panels (b–d), except that the leftmost panels show the AFM images with the mask regions highlighted in green.

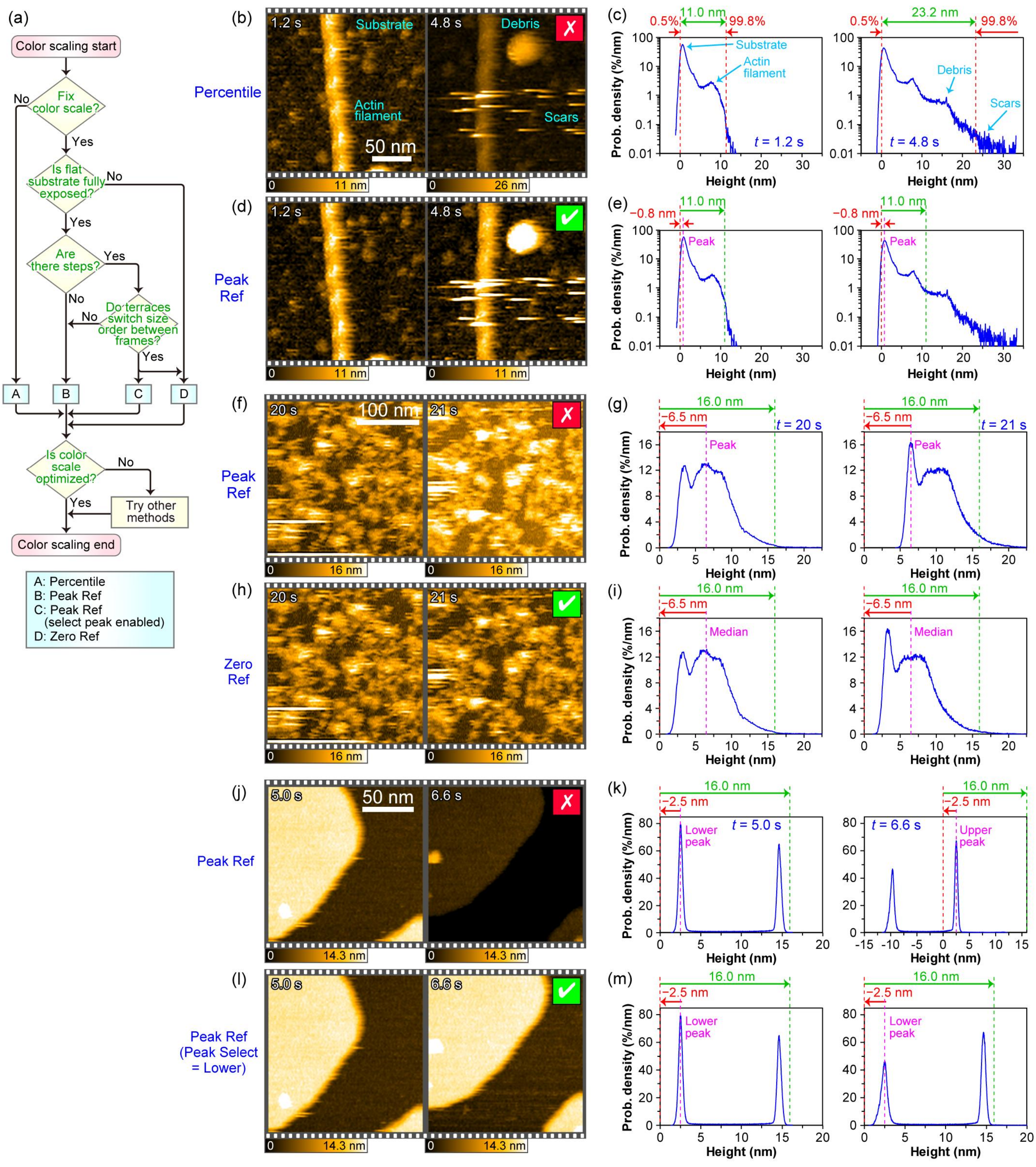

**Figure 4.** (**a**) Flowchart of the color-scaling procedure. (**b–e**) Comparison of the Percentile (b,c) and Peak Ref (d,e) color-scaling algorithms using HS-AFM images of an actin filament with a large portion of the substrate surface (b,d) and the corresponding height probability density distributions (c,e). (**f–i**) Comparison of the Peak Ref (f,g) and Zero Ref (h,i) color-scaling algorithms using

HS-AFM images fully covered by Smc5/6 complexes (f,h) and the corresponding height probability density distributions (g,i). (**j–m**) Comparison of the Peak Ref color-scaling algorithm without (j,k) and with Peak Select mode (l,m) using HS-AFM images exposing lipid membranes with step edges (j,l) and the corresponding height probability density distributions (k,m).

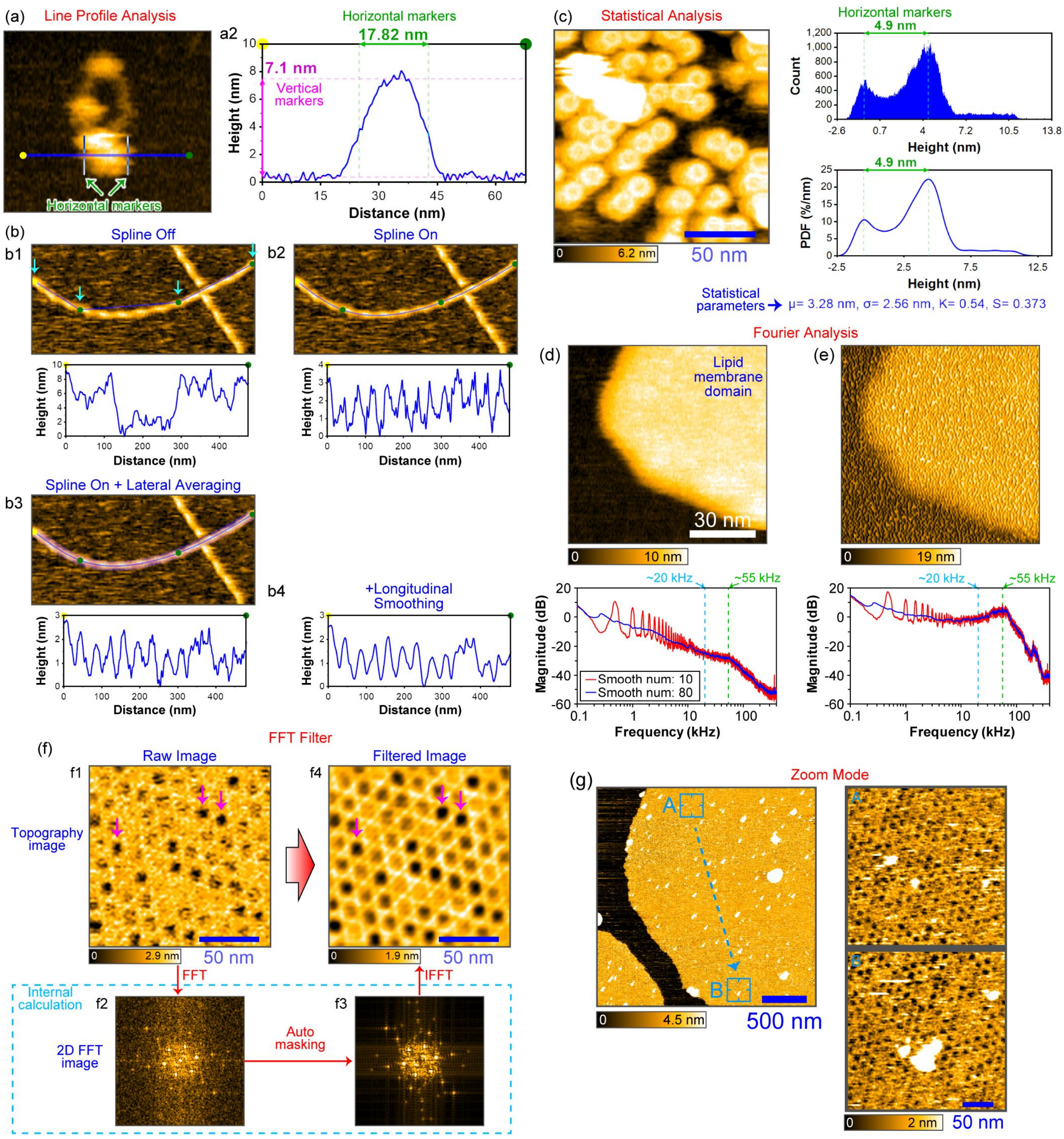

**Figure 5.** (**a**) Line-profile acquisition using a Smc5/6 complex: AFM image (a1) and corresponding profile (a2). The green and purple dashed lines indicate horizontal and vertical distance markers, with their distances shown above. (**b**) Effects of line profile settings using an actin filament: no optional settings (b1), spline interpolation (b2), lateral averaging (b3), and longitudinal smoothing (b4). Arrows in b1 indicate control points. (**c**) Height-distribution analysis of GroEL, showing the AFM image, height histogram, and KDE curve. (**d,e**) Fourier analysis of HS-AFM images of lipid membrane domain acquired under stable feedback (d) and feedback oscillation (e); the lower spectra show smoothed data used for bandwidth analysis, with moving-average coefficients of 18 and 80 for the red and blue spectra, respectively. (**f**) FFT filtering of a 2D Annexin V lattice: original topographic image (f1), 2D FFT magnitude image (f2), threshold-masked FFT magnitude image (f3), and inverse-FFT-filtered topographic image (f4). Purple arrows indicate honeycomb pit structures preserved after filtering. Fourier magnitude images are shown on a square-root scale. (**g**) Demonstration of the zoom mode. The left panel shows the original AFM topographic image, and the two panels on the right show magnified views of the regions indicated in the left image.

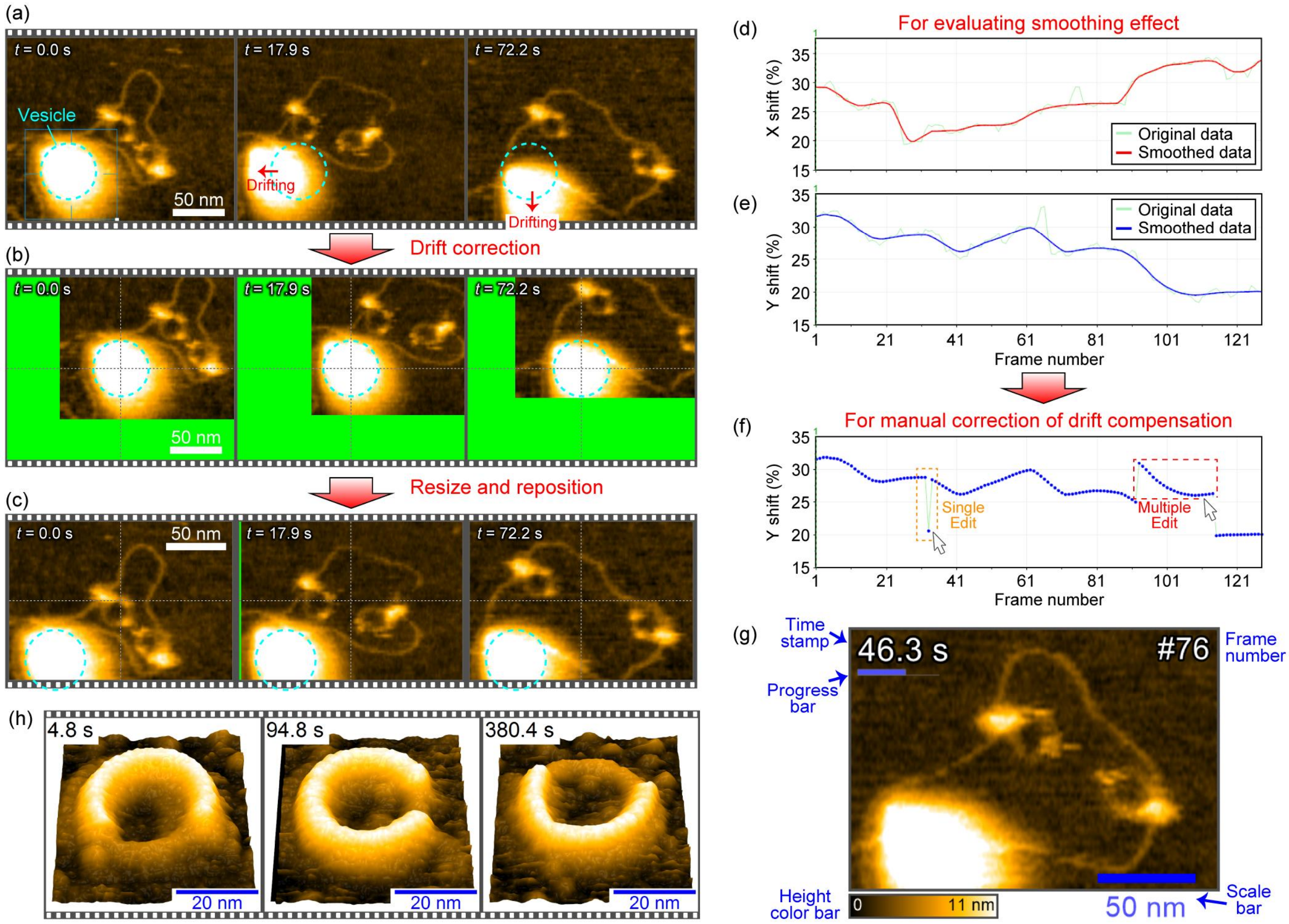

**Figure 6.** (**a–c**) Workflow of drift correction using HS-AFM images of Smc5/6 bound to DNA: original images (a), drift-corrected images (b), and images after size and position adjustment (c). In panel (a), the bounding box indicates the region selected in the software as the reference structure for cross-correlation-based alignment. (**d–f**) Time course of drift-correction shifts for evaluating the effect of smoothing in the X (d) and Y (e) directions and for mouse-based manual adjustment of individual points (f). The green regions indicate areas where no data are available. (**g**) Representative image and movie outputs from HS-AFM data, in which various types of information can be overlaid on the images. (**h**) 3D representation generated from the HS-AFM dataset of the MyD88 Toll/interleukin-1 receptor domain, originally published in Ref. [33]. The HS-AFM datasets shown in panels (a–c) and (g), as well as the final output movie, were previously reported in Ref. [62] and its Supplementary Material, and are reused here to demonstrate the analysis workflow.

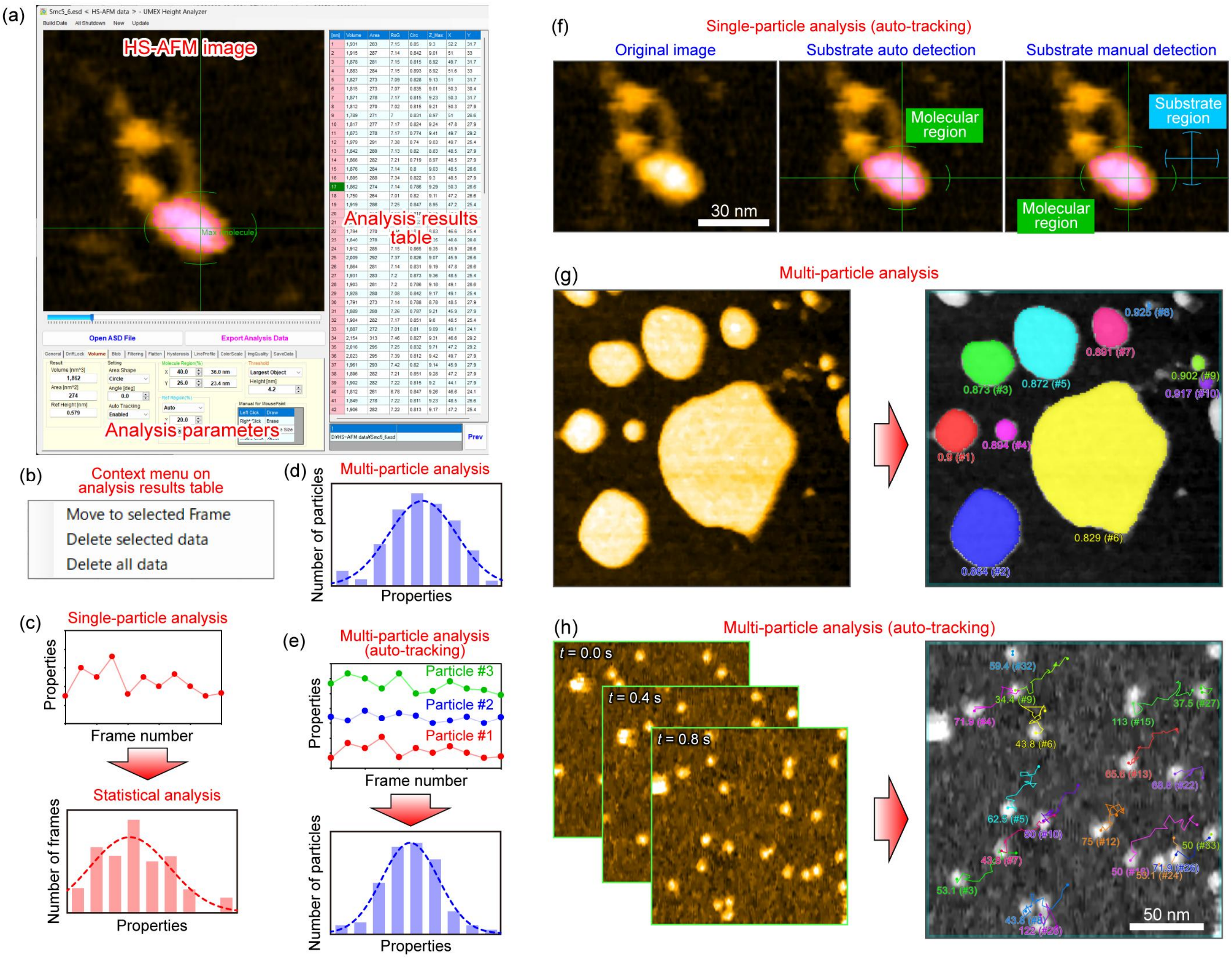

**Figure 7.** (**a**) Overall GUI of UMEX Height Analyzer. (**b**) Screenshot of the context menu for the analysis results table. (**c–e**) Schematic workflows of single-particle (c), multi-particle (d), and multi-particle auto-tracking (e) analyses. The quantified features are subsequently analyzed statistically using histograms. (**f–h**) Examples of single-particle analysis (Smc5/6; f), multi-particle analysis (lipid membrane islands on mica; g), and multi-particle trajectory analysis (streptavidin molecules on a lipid membrane; h). The original and analyzed AFM images are shown on the left and right, respectively.

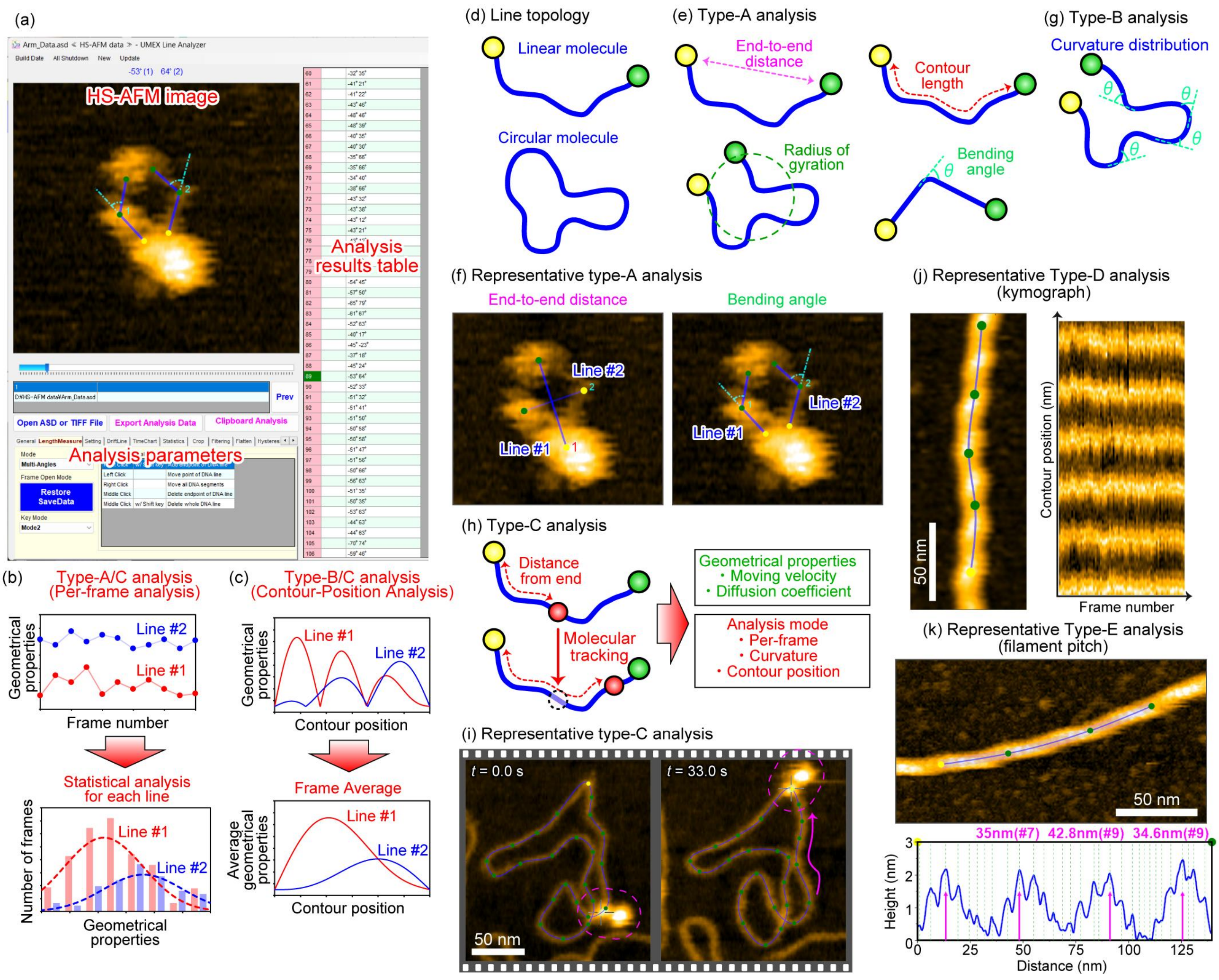

**Figure 8.** (**a**) Overall GUI of UMEX Line Analyzer. (**b,c**) Schematic workflows of Type-A/C (b) and Type-B/C (c) analyses. (**d**) Selectable line topologies for Type-A–C analyses. (**e,f**) Quantifiable geometrical parameters in Type-A analysis (e) and representative Type-A analysis of a Smc5/6 complex (f). (**g**) Quantifiable geometrical parameter in Type-B analysis. (**h,i**) Quantifiable geometrical parameters in Type-C analysis (h) and representative Type-C analysis of Smc5/6 bound to DNA (i). (**j,k**) Representative analyses of actin filaments using the Type-D (j) and Type-E (k) analysis modes. In the graph in panel (k), the purple arrows and green dashed lines indicate the major and minor pitches, respectively.

## Supporting Information

**Supplementary Figure 1:** File management and parameter-saving functions in UMEX Viewer.

**Supplementary Figure 2:** Flatten-masking workflow for HS-AFM images actin filaments.

**Supplementary Figure 3:** Workflow of manual mask editing for flattening.

**Movie S1:** HS-AFM videos demonstrating flattening without masking. Panels (a–c) correspond to the datasets shown in Fig. 3(b–d), respectively: XY line-by-line flattening applied to an isolated Smc5/6 complex (a) and clustered molecules (b), and plane flattening applied to a terrace step of a lipid membrane (c). From left to right, the videos show the ideally flattened image, and the flattening results obtained using the least-squares, Theil–Sen, and RANSAC algorithms.

**Movie S2:** Comparison of color-scaling algorithms. Panels (a, b) compare the Percentile and Peak Ref scaling using the actin-filament dataset in Fig. 4(b, d). Panels (c, d) compare the Peak Ref and Zero Ref scaling using the densely packed Smc5/6 dataset in Fig. 4(f, h). Panels (e, f) compare the Peak Ref scaling without and with peak-selection mode using the step-edged lipid-membrane dataset in Fig. 4(j, l).

**Movie S3:** HS-AFM videos demonstrating FFT filtering using the dataset of a 2D Annexin V crystal shown in Fig. 5(f): original HS-AFM image (a) and FFT-filtered HS-AFM image (b).

**Movie S4:** Demonstration of the zoom mode using the a 2D Annexin V crystal dataset shown in Fig. 5(g): original AFM image (a) and magnified view extracted from the original image (b).

**Movie S5:** Workflow of drift correction using the Smc5/6-bound DNA dataset shown in Fig. 6(a): original images (a), drift-corrected images (b), and size- and position-adjusted images (c).

**Movie S6:** Animated 3D visualization of HS-AFM data using the MyD88 Toll/interleukin-1 receptor domain dataset shown in Fig. 6(h).

**Movie S7:** Demonstration of single-particle trajectory analysis using the Smc5/6 complex dataset shown in Fig. 7(f). Panels (a) and (b) show the original and analyzed AFM images, respectively.

The intersection of the green crosshairs indicates the position of the maximum height and is automatically tracked from frame to frame.

**Movie S8:** Demonstration of multi-particle trajectory analysis using the dataset of streptavidin molecules on a lipid membrane shown in Fig. 7(h). Panels (a) and (b) show the original and analyzed AFM images, respectively.